\documentclass[Times,8pt,aps,prl,twocolumn,showpacs,amsmath,amssymb,floatfix,footinbib,superscriptaddress]{revtex4-1}
\usepackage{amsmath,natbib,graphicx,subfigure,amssymb,graphics,amsmath,mathrsfs,CJK,color}
\usepackage{multirow,fancyhdr,color,bm,tabularx,psfrag,geometry,dcolumn,datetime}
\usepackage[colorlinks=true,linkcolor=blue,urlcolor=blue,citecolor=blue]{hyperref}
\usepackage[mathlines]{lineno}%\pagewiselinenumbers
\def\footnoterule{\kern -1mm \hrule width 5.8cm \kern 2.2mm}%
\usepackage{tikz,xcolor,hyperref}% Make Orcid icon
\definecolor{lime}{HTML}{A6CE39}
\DeclareRobustCommand{\orcidicon}{%
 \begin{tikzpicture}
 \draw[lime, fill=lime] (0,0)
    circle [radius=0.16]
    node[white] {{\fontfamily{qag}\selectfont \tiny ID}};\draw[white, fill=white] (-0.0625,0.095)
    circle [radius=0.007];
 \end{tikzpicture}
\hspace{-2mm}}
\foreach \x in {A, ..., Z}
{\expandafter\xdef\csname orcid\x\endcsname{\noexpand\href{https://orcid.org/\csname orcidauthor\x\endcsname}{\noexpand\orcidicon}}}

\begin{document}

%\preprint{ }

\title{Dual peaks evoluting into single-peak for sub-wavelength 2-D atom localization in a V-type atomic system}%\LARGE\boldmath\bf

\thanks{Supported by the National Natural Science Foundation of China ( Grant Nos. 61205205 and 6156508508 ),
the General Program of Yunnan Provincial Research Foundation of Basic Research for application, China ( Grant No. 2016FB009 )
and the Foundation for Personnel training projects of Yunnan Province, China ( Grant No. KKSY201207068 ).}


\author{Shun-Cai Zhao\orcidA{}}
\email[Corresponding author: ]{ zhaosc@kmust.edu.cn }
\affiliation{Department of Physics, Faculty of Science, Kunming University of Science and Technology, Kunming, 650500, PR China}

\author{Xin Li}
\affiliation{Department of Physics, Faculty of Science, Kunming University of Science and Technology, Kunming, 650500, PR China}

\author{Ping Yang }
\affiliation{Department of Physics, Faculty of Science, Kunming University of Science and Technology, Kunming, 650500, PR China}

%\date{\today}

\begin{abstract}
The atom localization of a V-type atomic system is discussed by the detunings associated
with the probe and the two orthogonal standing-wave fields, and by the spontaneously
generated coherence (SGC). Within the half-wavelength domain in the 2-dimensional(2-D) plane,
the atom localization depicted by the probe dual absorption
peaks is achieved when the detunings are tuned. However, the dual peaks change into
single-peak when the SGC arises. The single-peak 2-D localization demonstrated the advantage
for atom localization achieved by the flexible manipulating parameters in our scheme.
\end{abstract}

%\pacs{ddfh  hhhh }
%\keywords{}


\maketitle
\section{INTRODUCTION}
%\linenumbers
Sub-wavelength atom localization becomes an
active research topic from both the theoretical and
experimental points of view\cite{1,2}.
Several schemes based on the optical methods are introduced to measure the position of the atom, such as the
spontaneous emission \cite{3,4,5}, absorption \cite{6,7},
population \cite{8,9,10,11,12}, the
entanglement between the atom`s internal state and its
position \cite{13}, the phase shift of the field\cite{14} and
gain \cite{15}.
In the schemes of an atom interacts with two orthogonal
standing-wave fields \cite{16,17,18,19}, the atomic position
information is given by multiple simultaneous quadrature measurement,
the population in the upper state or in any ground state measurement,
and incorporating the quantum interference phenomenon measurement,
etc..

However, the implemented atom localization is multi-peak probe absorbtion in most of the
optical methods\cite{6,7} or in the schemes of an atom interacts with two orthogonal
standing-wave fields\cite{16,17,18,19}. In our scheme, we analyzed the different parameters,
i.e., the SGC and the detunings associated with the probe and two orthogonal standing-wave fields.
The probe dual absorbtion peaks evoluting into single-peak for the atom localization via
the flexibly manipulating manners is achieved by both the external fields and
the quantum coherence, which demonstrates the advantage
over the dual peaks atom localization schemes\cite{16,17,18,19}.

\section{MODEL AND EQUATION}

Let us consider a V-type system[see Fig.1] with the lower state
$|1\rangle$ and two excited states $|2\rangle$ and $|3\rangle$. The
transition $|1\rangle$$\leftrightarrow$$|3\rangle$ with frequency
$\omega_{13}$ is driven by a weak probe field of frequency
$\nu_{p}$ with Rabi frequency $\Omega_{p}$ being
$\Omega_{p}$=$E_{p}\mu_{13}/2\hbar$. A strong coherent coupling
standing-wave field $E_{x,y}$ with the same frequency $\nu_{c}$ is
simultaneously applied to the transition
$|1\rangle$$\leftrightarrow$$|2\rangle$ with frequency
$\omega_{12}$. The Rabi frequency $\Omega_{c}(x,y)$ corresponding to
the composition of two orthogonal standing waves is
$\Omega_{c}(x,y)$=$E_{x,y}\mu_{12}/2\hbar$=$\Omega_{0}[sin(\kappa_{1}x+\delta)+sin(\kappa_{2}y+\eta)]$.
Where $\mu_{13}$ and $\mu_{12}$ are the corresponding dipole matrix
elements, $\kappa_{i}=2\pi/\lambda_{i},(i=1,2)$ is the wave vector
with wavelengths $\lambda_{i},(i$=$1,2)$ of the corresponding standing
wave fields and the parameters $\delta$ and $\eta$ are the phase shifts associated
with the standing-wave fields. $\Delta_{p}$=$\omega_{13}-\nu_{p}$, and
$\Delta_{c}$=$\omega_{12}-\nu_{c}$ are the detunings of the two
corresponding fields, respectively. For simplicity, we assume
$\Omega_{p}$ and $\Omega_{0}$ to be real. $2\gamma_{1}$ and
$2\gamma_{2}$ are the spontaneous decay rates of the excited states
$|2\rangle$ and $|3\rangle$ to the ground states $|1\rangle$,
respectively. And in the atomic system, the two excited levels $|2\rangle$ and
$|3\rangle$ are closely spaced such that the two transitions to the
lower state interact with the same vacuum mode, SGC can be present.
Which was proposed the realization in photonic band-gap materials
and semiconductor quantum dots\cite{20}.

\begin{figure}[!h]
\centering\includegraphics[width=2.5in]{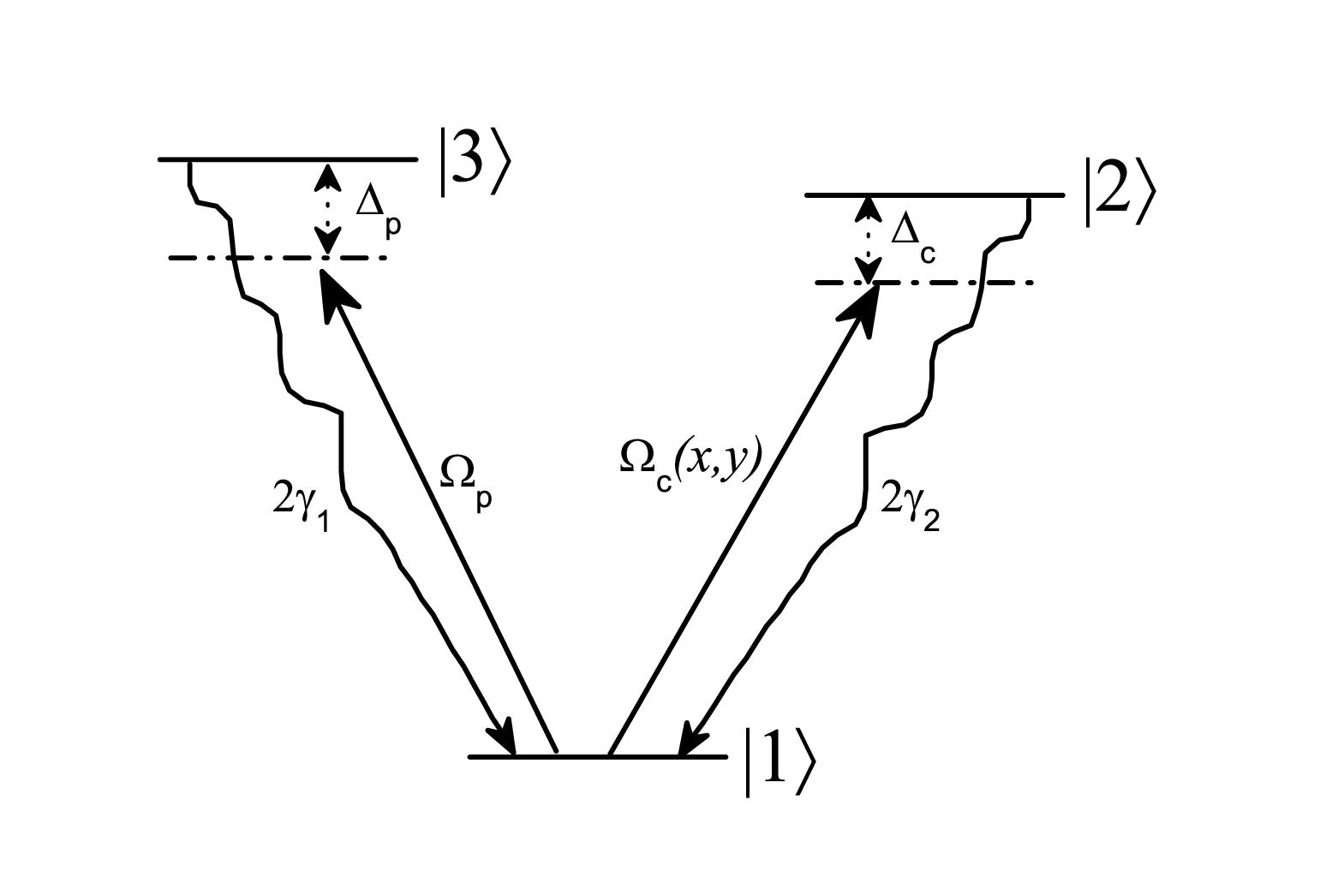}
\caption{Schematic diagrams. The position-dependent Rabi frequency
$\Omega_{c}(x,y)$ corresponding to the atomic transition from
$|2\rangle$ to $|1\rangle$. The transition from $|1\rangle$ to $|3\rangle$ is coupled
via a weak probe field $\Omega_{p}$. $2\gamma_{1}$ and $2\gamma_{2}$
are the atomic decay rates.}
\label{fig_1}
\end{figure}

\noindent Here, the center-of-mass position distribution of the atom is assumed
to be nearly uniform along the direction of the standing wave. Hence,
we can apply the Raman-Nath approximation and ignore the kinetic
energy of the atom in the Hamiltonian\cite{21}. Then, under the electric
dipole and rotating-wave approximation, the systematic density
matrix in the interaction picture involving SGC can be written
as\cite{22},
\vskip -0.5cm
\begin{align}
\dot{\rho}_{22}&=-2\gamma_{1}\rho_{22}+i\Omega_{c}(\rho_{12}-\rho_{21})-p\sqrt{\gamma_{1}\gamma_{2}}(\rho_{23}+\rho_{32}),\nonumber\\
\dot{\rho}_{33}&=-2\gamma_{1}\rho_{33}+i\Omega_{p}(\rho_{13}-\rho_{31})-p\sqrt{\gamma_{1}\gamma_{2}}(\rho_{23}+\rho_{32}),\nonumber\\
\dot{\rho}_{12}&=-(\gamma_{1}+i\Delta_{c})\rho_{12}-p\sqrt{\gamma_{1}\gamma_{2}}\rho_{13}+i\Omega_{c}(\rho_{22}-\rho_{11})\nonumber\\
                &   +i\Omega_{p}\rho_{32}, \\
\dot{\rho}_{13}&=-p\sqrt{\gamma_{1}\gamma_{2}}\rho_{12}-(\gamma_{2}+i\Delta_{p})\rho_{13}+i\Omega_{c}\rho_{23}+i\Omega_{p}\nonumber\\
                &   (\rho_{33}-\rho_{11}), \nonumber\\
\dot{\rho}_{23}&=i\Omega_{c}\rho_{13}-i\Omega_{p}\rho_{21}-i(\Delta_{p}-\Delta_{c})\rho_{23}-(\gamma_{1}+\gamma_{2})\rho_{23}\nonumber\\
                &-p\sqrt{\gamma_{1}\gamma_{2}}(\rho_{22}+\rho_{33}).\nonumber
\end{align}
\noindent The atom is initially in the ground state, therefore,
we use the initial condition \(\rho^{0}_{11}=1\), \(\rho^{0}_{22}=0\), \(\rho^{0}_{33}=0\)
along with $\rho_{ij}^{\ast}=\rho_{ji}$.
Here, the parameter p denotes the alignment of the two dipole
moments and is defined as
$p=\vec{\mu_{13}}\cdot\vec{\mu_{12}}/|\vec{\mu_{13}}\cdot\vec{\mu_{12}}|$=$cos\theta$
with $\theta$ being the angle between the two dipole moments
$\vec{\mu_{12}}$ and $\vec{\mu_{13}}$, which is very sensitive for the
existence of SGC effect, and can be a random angle between 0 and
2$\pi$ except for 0 and $\pi$. The terms with
$p\sqrt{\gamma_{1}\gamma_{2}}$ represent the quantum interference
resulting from the cross coupling between spontaneous emission paths
$|1\rangle$-$|2\rangle$and $|1\rangle$-$|3\rangle$. It should be
noted that only for small energy spacing between the two excited
levels are the interference terms in the systematic density matrix
significant; otherwise the oscillatory terms will average out to
zero and thereby SGC effect vanishes.

Because of the spatial-dependent atom-field interaction, 2-D atom
localization can be realized via measuring the probe gain or
absorption\cite{15}. The information about the atomic position from
the susceptibility\cite{19} at the probe field frequency is what we want to get,
and the nonlinear Raman susceptibility $\chi$ is then given by

\begin{equation}\label{2}
\begin{split}
\chi=\frac{2N|\mu_{13}|^{2}}{\epsilon_{0}\Omega_{p}\hbar}\rho_{13},
\end{split}
\end{equation}

\noindent where N is the atom number density in the medium and $\mu_{13}$ is
the magnitude of the dipole-matrix element between $|1\rangle$ and
$|3\rangle$. $\epsilon_{0}$ is the permittivity in free space.
When the probe field is weak, i.e. \(\Omega_{p}\ll\Omega_{c}\),
it is sufficient to keep only the linear terms of Rabi frequency
\(\Omega_{p}\) in equations (1). By use of these approximations,
we find the first order steady-state analytical solution for $\rho_{13}$ can be
written as
\begin{equation}\label{3}
\begin{split}
\rho^{1}_{13}=&\frac{-4A_{7}+\Omega_{p}\{-A_{1}\rho^{0}_{12}+\Omega_{c}[-i(A_{0}-A_{5})p\rho^{0}_{31}}{4[-p^{2}A_{8}+A_{10}p^{4}-4p^{6}} \\
              &\frac{-2A_{2}p\rho^{0}_{13}+2A_{3}\rho^{0}_{21}+2pA_{6}\Omega_{c}\rho^{0}_{23}]-2pA_{4}\rho^{0}_{32}\}}{+A_{9}(1+\Delta_{c}^{2}+2\Omega_{c}^{2})]}
\end{split}
\end{equation}
where the matrix elements \(\rho^{0}_{12}\), \(\rho^{0}_{13}\) and \(\rho^{0}_{23}\) are the zero order
steady-state analytical solutions,
\begin{align*}
\rho^{0}_{12}&=\frac{(-i+\Delta_{p})(-2i-\Delta_{c}+\Delta_{p})\Omega_{c}}{(2i+\Delta_{c}-\Delta_{p})[p^{2}+(\Delta_{p}-i)(\Delta_{c}-i)]}\\
               &\frac{-\Omega_{c}^{3}}{+(\Delta_{c}-i)\Omega_{c}^{2}},\\
\rho^{0}_{13}&=\frac{p\Omega_{c}(2i+\Delta_{c}-\Delta_{p})}{(2i+\Delta_{c}-\Delta_{p})[ip^{2}-i+\Delta_{c}(1+i\Delta_{p})+\Delta_{p}]} \\
               &\frac{}{+(1+i\Delta_{c})\Omega_{c}^{2}},               \\
\rho^{0}_{23}&=\frac{ip\Omega_{c}^{2}}{(2i+\Delta_{c}-\Delta_{p})[p^{2}+(-i+\Delta_{p})(-i+\Delta_{c})]+(-i}\\
               &\frac{}{+\Delta_{c})\Omega_{c}^{2}},
 \end{align*}
\noindent where the parameters $A_{i},(i=0,\ldots,10)$ in Eq.(3) are given in
the Appendix. $\gamma_{1}=\gamma_{2}=\gamma$ and all the parameters
are reduced to dimensionless units by scaling with $\gamma $. Using
Eq.(2), which consists of both real and imaginary
parts, i.e., $\chi=\chi^{'} + i\chi^{''}$. The imaginary part of the
susceptibility gives the absorption profile of the probe field which
can be written as

\begin{equation}\label{4}
\begin{split}
\chi^{''}=\frac{2N|\mu_{13}|^{2}}{\epsilon_{0}\hbar}Im[\frac{\rho_{13}}{\Omega_{p}}]=\alpha
Im[\frac{\rho_{13}}{\Omega_{p}}],
\end{split}
\end{equation}

\noindent where $\alpha=\frac{2N|\mu_{13}|^{2}}{\epsilon_{0}\hbar}$. Here we
are interested in the position information of the atom via the
absorption process of the probe field. Then Eq.(4) is the main result
and reflects the atomic position probability distribution \cite{6,7}. Therefore,
we can obtain the atomic position information via the corresponding parameters, i.e.,
the intensities of SGC and the detunings associated with the probe and two orthogonal
standing-wave fields.

\section{Numerical results and discussions}
\subsection{2-D atom localization via the detuning $\Delta_{p}$}
In this following, we analyze the position information via
the imaginary part $\chi^{''}$ of the susceptibility \cite{6,7}. However,
the analytical expression for the position probability $\chi^{''}$
in Eq.(4) corresponding to the field detunings and intensities of SGC is rather
cumbersome. Hence, we follow the numerical approach to get the
position distribution information.

\begin{figure}[!ht]
\centering\includegraphics[width=1.5in]{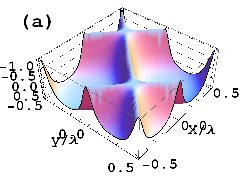}\includegraphics[width=1.5in]{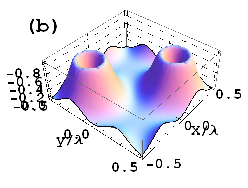}
\centering\includegraphics[width=1.5in]{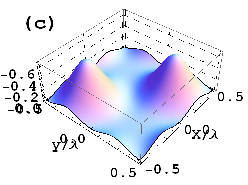}\includegraphics[width=1.5in]{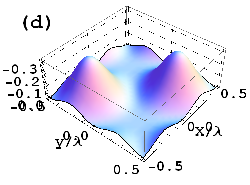}
\caption{(Color online) Plots for 2-D atom localization: Im[$\chi$] as a
function of (x,y) versus the probe field detuning
$\Delta_{p}$. (a)$\Delta_{p}$=0, (b)$\Delta_{p}$=20$\gamma$,
(c)$\Delta_{p}$=30$\gamma$, (d)$\Delta_{p}$=40$\gamma$. The other
parameters used are $\kappa_{1}$=$\kappa_{2}$=$2\pi$/$\lambda$,
$\Omega_{c0}$=10$\gamma$, $\Omega_{p}$=0.01$\gamma$,
$\delta$=$\eta$=0, $\theta$=0.5$\pi$, and where $\gamma$ is the
scaling parameter.}
\label{fig_2}
\end{figure}

\vskip 2mm
\begin{figure}[!ht]
\centering\includegraphics[width=1.5in]{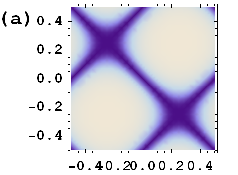}\includegraphics[width=1.5in]{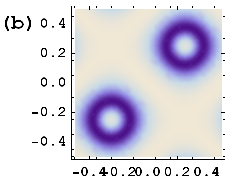}
\centering\includegraphics[width=1.5in]{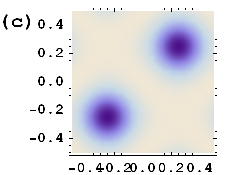}\includegraphics[width=1.5in]{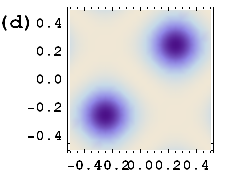}
\caption{(Color online) Density plots of 2-D atom localization:
Im[$\chi$] as a function of (x,y) versus the probe field
detuning $\Delta_{p}$. All other parameters in (a) to (d) are the same
as in Figs.2(a) to 2(d), respectively.}
\label{fig_3}
\end{figure}

\noindent Initially, we consider the position information dependent the probe field detuning $\Delta_{p}$.
The standing-wave field is tuned to interact with the transition $|1\rangle$$\leftrightarrow$$|2\rangle$ resonantly,
and $\Omega_{c0}$$=$$10\gamma$, $\Omega_{p}$$=$$0.01\gamma$. The position-dependent the probe detuning
$\Delta_{p}$ is shown in Fig.2 (a)$\Delta_{p}$=0, (b)$\Delta_{p}$=20$\gamma$, (c)$\Delta_{p}$=30$\gamma$,
(d)$\Delta_{p}$=40$\gamma$. And their density plots in Fig.2 are shown in Fig.3.
As noted in Fig.2(a), the equal localization peaks distribute on the
diagonal in the second and fourth quadrants within the half wavelength domain
[see Figs. 2(a) and 3(a)]. As the detuning $\Delta_{p}$ increasing, we observe the
position distribution of the atom is situated in the first
and third quadrants with a craterlike pattern [see Figs. 2(b)],
and the atom is localized at the circular edges of the craters [see Fig. 3(b)]
. Moreover, when $\Delta_{p}$ is tuned to an appropriate value [e.g., 30$\gamma$, 40$\gamma$ in Fig. 2(c) and (d)],
the position distribution displays two equivalent spikelike pattern in the quadrants I and
III[see in Fig. 2(c) and (d)]. Which shows that the spatial resolution is improved via the light spots in Fig.3.
These results demonstrate the increasing resolution for 2-D atom localization can be obtained by the proe detuning $\Delta_{p}$.

\subsection{2-D atom localization via the detuning $\Delta_{c}$}

\vskip 2mm
\begin{figure}[!ht]
\centering\includegraphics[width=1.5in]{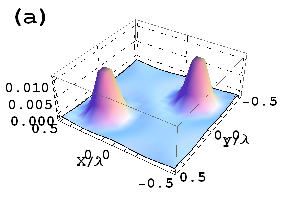}\includegraphics[width=1.5in]{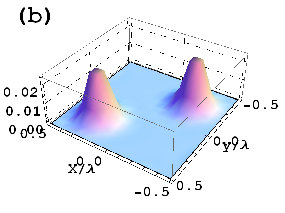}
\centering\includegraphics[width=1.5in]{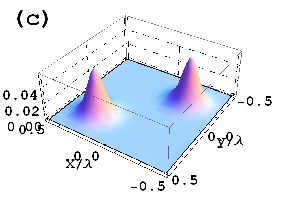}\includegraphics[width=1.5in]{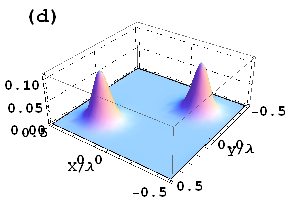}
\caption{(Color online) Plots for 2-D atom localization: Im[$\chi$] as a
function of (x,y) versus the coupled standing-wave field
detuning $\Delta_{c}$. (a)$\Delta_{c}$=8$\gamma$,
(b)$\Delta_{c}$=12$\gamma$, (c)$\Delta_{c}$=15$\gamma$,
(d)$\Delta_{c}$=20$\gamma$, $\Delta_{p}$=20$\gamma$. All other
parameters are the same as in Fig.2.}
\label{fig_4}
\end{figure}

\vskip 2mm
\begin{figure}[!ht]
\centering\includegraphics[width=1.5in]{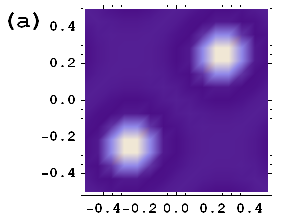}\includegraphics[width=1.5in]{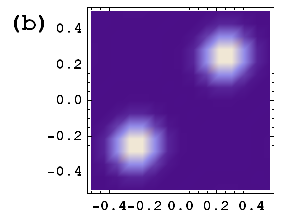}
\centering\includegraphics[width=1.5in]{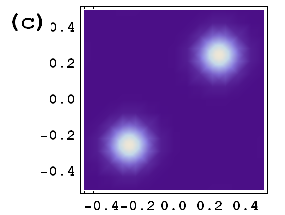}\includegraphics[width=1.5in]{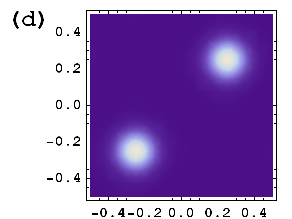}
\caption{(Color online) Density plots of 2-D atom localization:
Im[$\chi$] as a function of (x,y) versus the probe field
detuning $\Delta_{c}$. All other parameters in (a) to (d) are the same
as in Figs.4(a) to 4(d), respectively.}
\label{fig_5}
\end{figure}

Secondly, we consider the effect of the detuning $\Delta_{c}$ on the 2-D
atom localization. Although the values of some parameters are the same as
in Fig. 2, the behavior of the atom localization in Fig.4 is different from those in the
previous subsection. Similarly, the corresponding density plot of Fig.4 is
given in Fig.5. When the detuning of two orthogonal standing-wave fields is set as
8$\gamma$ in Figs.4 (a), 12$\gamma$ in Figs.4(b), 15$\gamma$ in Figs.4(c) and 20$\gamma$ in Figs.4(d)
with the probe field coupling the atomic system off-resonantly, i.e. $\Delta_{p}$=20$\gamma$,
two spikelike patterns with equal increasing peak maxima occur in the quadrants I and III of the
x-y plane. And their peak values increase from 0.01 to 0.1. Meanwhile, the spikelike structure become more and more acuity,
and the dual light-spot size in Fig.5 confirms these straightly. As a result, we can achieve an increasing precision
resolution for 2-D atom localization by measuring the detuning $\Delta_{c}$ under
the condition of the probe field coupling the atomic system off-resonance.
Then an increasing resolution for 2-D atom localization can be obtained by the above two different detunings.

\subsection{2-D atom localization via the intensities of SGC }

\vskip 2mm
\begin{figure}[!ht]
\centering\includegraphics[width=1.5in]{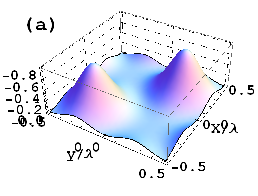}\includegraphics[width=1.5in]{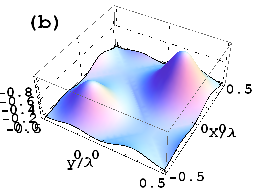}
\centering\includegraphics[width=1.5in]{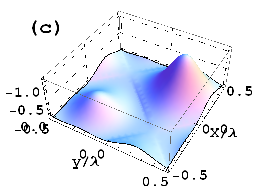}\includegraphics[width=1.5in]{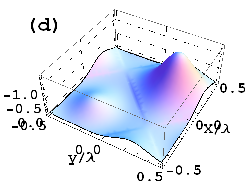}
\caption{(Color online) Plots for 2-D atom localization: Im[$\chi$] as a
function of (x,y) versus the intensities p of SGC.
(a)$\theta=\pi/1.99$, (b)$\theta=\pi/1.9$, (c)$\theta=\pi/1.8$,
(d)$\theta=\pi/1.7$. $\Delta_{c}$=0, $\Delta_{p}$=30$\gamma$. All
 other parameters are the same as in Fig.2.}
\label{fig_6}
\end{figure}

As a matter of fact, we are particularly interested in
the influence of the intensities p=$cos\theta$ of SGC on the behavior of 2-D atom
localization, we give the 2-D localization dependent the intensities p=$cos\theta$
of SGC, where $cos\theta$ depicting the intensities of SGC is set as
(a)$\theta$=$\pi/1.99$, (b)$\theta$=$\pi/1.9$, (c)$\theta$=$\pi/1.8$, (d)$\theta$=$\pi/1.7$ in Fig.6.
The corresponding density plots are illustrated in Fig. 7. For the sake of simplification,
we set the standing-wave field drives the transition$|1\rangle$$\leftrightarrow$$|2\rangle$
resonantly, i.e, $\Delta_{c}$=0, and $\Delta_{p}$=30$\gamma$. All other parameters
are the same as in Fig.2.

\vskip 2mm
\begin{figure}[!ht]
\centering\includegraphics[width=1.5in]{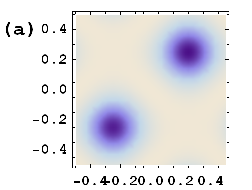}\includegraphics[width=1.5in]{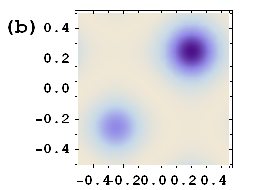}
\centering\includegraphics[width=1.5in]{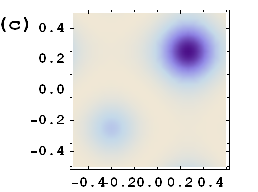}\includegraphics[width=1.5in]{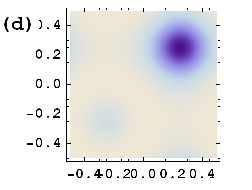}
\caption{Color online) Density plots of 2-D atom localization:
Im[$\chi$] as a function of (x,y) versus the intensities p
of SGC. All other parameters in (a) to (d) are the same as in
Figs.6(a) to 6(d), respectively.}
\label{fig_7}
\end{figure}

As noted In Fig.6(a), two spikelike patterns with the equal peak
maxima(-0.8) appear in the first and third quadrants[also see Figs.7(a)].
When the $\theta$ corresponding to the intensity of SGC is detected at $\pi/1.9$, the peak
maxima in quadrant III decreases while the other one in quadrant I remains, which shows
the probability of the atom distributing in the third quadrant decreases [ see Figs.6(b) and 7(b)].
For the case of $\theta$=$\pi/1.8$,
the single-spikelike pattern is distributed in the first quadrant
and very few in the third quadrant, as can be seen Figs.6(c) and Figs.7(c).
Continuing to adjust $\theta$=$\pi/1.7$, there's almost zero peak distribution in the third quadrant [ see Figs.6(d) and 7(d)],
which demonstrates the spatial resolution of atomic position is greatly improved.
The identical multi-peak scheme for 2-D localization has the
disadvantage that it is unknown which one is responsible for a possible atomic position distribution.
However, one can determine more precisely where the atoms is via the single-peak
position distribution. So, a better resolution can achieved by the tuned intensities p
of SGC than by the previous detunings.

\section{CONCLUSION}

In summary, the roles of SGC and detunings associated with the probe
and two orthogonal standing-wave fields in the 2-D atom localization were investigated in our scheme.
Two equal and flexibly tunable localization peaks for the 2-D atom localization are observed when the
detunings were tuned. However, when the intensities of SGC were manipulated,
two peaks with one increasing and other decreasing are observed, and finally the single-peak is shown in the x-y plane.
The single-peak for 2-D atom localization has more advantage than the multi-peak case
because of the determining atomic position more precisely via the single localization peak.
So, the advantage in 2-D atom localization via both the external
fields and the quantum coherence is obtained in our scheme.

\appendix*\section{appendix}
The parameters $A_{i}$ in Eq.(3) are given by the following:

\begin{align*}
A_{0}&=P^{4}(8i+\Delta_{c}-3\Delta_{p})+\Omega_{c}^{2}(5i+\Delta_{c}-2\Delta_{p})[\Omega_{c}^{2}\\
     &+(i+\Delta_{p})(-2i+\Delta_{c}-\Delta_{p})]+P^{2}[(i+\Delta_{p}) \\
     & (i+\Delta_{c})(8i+\Delta_{c}-3\Delta_{p})+2\Omega_{c}^{2}(5i+\Delta_{c}-4\Delta_{p})],  \\
A_{1}&=2iP^{2}\{(8i+\Delta_{c}-3\Delta_{p})[P^{2}+(i+\Delta_{p})(i+\Delta_{c})]+\Omega_{c}^{2}\\
     &(4i+\Delta_{c}-3\Delta_{p})\},\\
A_{2}&=2P^{4}-[\Omega_{c}^{2}+(i+\Delta_{p})(-2i+\Delta_{c}-\Delta_{p})]\\
      &[\Omega_{c}^{2}+(i+\Delta_{c})(-3i+\Delta_{p})]+ P^{2}[4+\Omega_{c}^{2}+\\
      &\Delta_{p}(i+\Delta_{p})+\Delta_{c}(5i+\Delta_{p})],\\
A_{3}&=-iP^{4}(-4i+\Delta_{c}-3\Delta_{p})+2[\Omega_{c}^{2}+(i+\Delta_{p})(-2i+\Delta_{c}\\
     &-\Delta_{p})](1+\Delta_{c}^{2}+2\Omega_{c}^{2})-iP^{2}\{\Delta_{p}(5-3i\Delta_{p})\\
     &+\Delta_{c}^{2}(3i+\Delta_{p})+\Omega_{c}^{2}(2i-3\Delta_{p})+\Delta_{c}[9-\Delta_{p}\\
     &(8i+3\Delta_{p})+\Omega_{c}^{2}] \},\\
A_{4}&=8iP^{4}+P^{2}\{-2i[8+\Delta_{c}(-4i+\Delta_{c})-4i\Delta_{p}-6\Delta_{c}\Delta_{p}\\
      &+\Delta_{p}^{2}]+\Omega_{c}^{2}(6i+\Delta_{c}-5\Delta_{p})\}+[\Omega_{c}^{2}\\
       & +(i+\Delta_{p})(-2i+\Delta_{c}-\Delta_{p})][2(1-i\Delta_{c})(2i+\Delta_{c}-\Delta_{p})\\
       &+\Omega_{c}^{2}(3i+\Delta_{c}-2\Delta_{p})],\\
\end{align*}
\begin{align*}
A_{5}&=P^{4}(4i+\Delta_{c}-3\Delta_{p})+[\Omega_{c}^{2}+(i+\Delta_{p})(-2i+\Delta_{c}-\Delta_{p})]\\
       &[2(i+\Delta_{c})(3+i\Delta_{p})+\Omega_{c}^{2}(7i+\Delta_{c}-2\Delta_{p})]+P^{2}\{\Delta_{c}^{2}\\
       &(i+\Delta_{p})+\Delta_{p}(-3-5i\Delta_{p}-8\Omega_{c}^{2})+8i(-2+\Omega_{c}^{2})+\Delta_{c}\\
       &[1+(4i-3\Delta_{p})\Delta_{p}+2\Omega_{c}^{2}]\},\\
A_{6}&=P^{2}(6i+\Delta_{c}-5\Delta_{p})+(5i+\Delta_{c}-2\Delta_{p})[(-2i+\Delta_{c}\\
      &-\Delta_{p})(i+\Delta_{p})+\Omega_{c}^{2}],\\
A_{7}&=-4iP^{5}\Omega_{c}+iP^{3}[8+\Delta_{c}^{2}+\Delta_{p}(-4i+\Delta_{p})-2\Delta_{c}(2i+3\Delta_{p})]\\
       &\Omega_{c}+p\Omega_{c}[(2+i\Delta_{c}-i\Delta_{p}+i\Omega_{c}^{2})(i+\Delta_{p})][(i+\Delta_{c})\\
       &(2i+\Delta_{c}-\Delta_{p})+\Omega_{c}^{2}]+(1+\Delta_{c}^{2}+2\Omega_{c}^{2})+\{[4+(\Delta_{c}\\
       &-\Delta_{p})^{2}](i+\Delta_{p})+(2i+\Delta_{c}-\Delta_{p})\Omega_{c}^{2}\}\Omega_{p}+P^{2}\\
       &\{(\Delta_{c}-i)[8+\Delta_{c})^{2}+\Delta_{p}(\Delta_{p}-4i)-2\Delta_{c}(2i+3\Delta_{p})]\\
       &-[12i+5\Delta_{c}+(3+2i\Delta_{p})\Delta_{p}]\Omega_{c}^{2}- i\Omega_{c}^{4}\}\Omega_{p}\\
       &+2P^{4}[-2\Delta_{c}+i(2+\Omega_{c}^{2})]\Omega_{p},\\
A_{8}&=-2\Delta_{c}^{3}\Delta_{p}+6(2+\Delta_{p})^{2}-2\Delta_{c}\Delta_{p}(6+\Delta_{p})^{2}\\
      &+\Delta_{c}^{2}(6+8\Delta_{p})^{2}+2[4-(\Delta_{c}-4\Delta_{p})(\Delta_{c}+\Delta_{p})]\\
      &\Omega_{c}^{2}+\Omega_{c}^{4}, \\
A_{9}&=[4+(\Delta_{c}-\Delta_{p})^{2}](1+\Delta_{p})^{2})+2[2+(\Delta_{c}-\Delta_{p})\Delta_{p}]\\
      &\Omega_{c}^{2}+\Omega_{c}^{4}, \\
A_{10}&=12+\Delta_{c}^{2}-10\Delta_{c}\Delta_{p}+\Delta_{p}^{2}-4\Omega_{c}^{2}.
\end{align*}

%\begin{widetext}
%\end{widetext}

% \begin{figure}
% \end{figure}

% Use \appendix* if there
%\appendix
%\section{}

%\begin{acknowledgments}
%\end{acknowledgments}


\end{document}